\documentstyle[aas2pp4,epsf]{article}

\received{24 June 1996}
\accepted{December 1996}

\slugcomment{Submitted to the {\it Astronomical Journal}}

\lefthead{F. La Franca and S. Cristiani}
\righthead{The QSO evolution from the {\it HBQS}}

\begin{document}

\title{The QSO evolution derived from the {\it HBQS} \\ 
and other complete QSO surveys
\footnote{Based on data of the ESO Key-Programme ``A Homogeneous Bright
QSO Survey''.}}

\author{Fabio La Franca}
\affil{Dipartimento di Fisica, Universit\`a degli studi ``Roma TRE''\\
Via della
Vasca Navale 84, I-00146 Roma, Italy\\
{\rm Electronic mail: lafranca@astrpd.pd.astro.it}}

\and
\author{Stefano Cristiani}
\affil{Dipartimento di Astronomia, Universit\`a degli studi di Padova \\
Vicolo dell'Osservatorio 5, I-35122 Padova, Italy\\
{\rm Electronic mail: cristiani@astrpd.pd.astro.it}\\ $~~~~$\\
$~~~~$\\$~~~~$\\
{\rm Submitted to the} The Astronomical Journal
}

\begin{abstract}
An Homogeneous Bright QSO Survey (HBQS) has been carried out in the framework
of an ESO Key programme. 327 QSOs (with $M_B<-23$ and $0.3<z<2.2$) have been
selected over an area of 555 $deg^2$ in the magnitude range $15<B<18.75$.  For
magnitudes brighter than $B=16.4$ the QSO surface density turns out to be a
factor 2.2 higher than what measured by the PG survey, corresponding to a
surface density of $0.013^{+.007}_{-.005} ~ deg^{-2}$. If the data from the
Edinburgh QSO Survey are included, an overdensity of a factor 2.7 is observed,
corresponding to a surface density of $0.016\pm 0.005 ~ deg^{-2}$. In order to
derive the QSO optical luminosity function we used Monte Carlo simulations that
take into account of the selection criteria, photometric errors and QSO
spectral slope distribution. We have combined our data with the Edinburgh QSO
Survey, the Large Bright QSO Survey, Durham/AAT survey, ESO/AAT faint survey
and the (ZM)$^2$B survey. 

The luminosity function can be represented with a Pure Luminosity Evolution,
$L(z)\propto (1+z)^k$, of a two power-law both for $q_0 = 0.5$ and $q_0 =
0.1$. For $q_0 = 0.5$ the $k$ evolution parameter is $k=3.26\pm 0.07$, slower
than the previous Boyle's (1992) estimate $k=3.45$. A flatter slope $\beta =
-3.72 \pm 0.13$ of the bright part of the luminosity function is also required.
If a spread in the QSO spectral slope of $\sigma_\gamma = 0.3$ and $0.5$ is
taken into account, the $k$ parameter drops to $3.2$ and $3.0$ respectively.
The observed overdensity of bright QSOs is concentrated at redshifts lower than
0.6. It results that in the range $0.3<z<0.6$ the luminosity function is
flatter than observed at higher redshifts. In this redshift range, for
$M_B<-25$, $32$ QSOs are observed instead of $19$ expected from our best-fit
PLE model. This feature requires a luminosity dependent luminosity evolution in
order to adequately represent the data in the whole $0.3<z<2.2$ redshift
interval. The observed overdensity of low redshift bright QSOs could be
originated by the contribution of a slower evolving population of radio-loud
QSOs. 

\end{abstract}


\section{INTRODUCTION}

The study of the evolution of the QSO Luminosity Function (LF) is of primary
importance in order to understand the physic driving the central "engine" of
AGNs (e.g. Efstathiou \& Rees 1988, Haehnelt \& Rees 1993). Studies of the
large scale structure of the Universe (e.g. Andreani \& Cristiani 1993), and of
the extragalactic X-ray and UV background (e.g. Madau, Ghisellini \& Fabian
1994; Comastri {\it et al.} 1995; Haardt \& Madau 1996) are based on the
knowledge of the optical QSO LF and  its relationship with other wavebands. The
LF of the optically selected QSOs is also the starting point  to disentangle
the evolutionary properties of the two populations of the radio-loud and
radio-quiet QSOs (La Franca {\it et al.} 1994). 

The general behaviour of the LF  is fairly well established in the redshift
interval $0.3<z<2.2$ for which color techniques provide reliable selection
methods (see Hartwick and Schade (1990), Boyle (1992) and Hewett and Foltz
(1994) for a review of the subject). The prevailing model was proposed by
Boyle, Shanks \& Peterson (1988) and was based on a sample of 420 faint
($b<20.9$) AGNs, combined with the Palomar Green (PG) Bright QSO Sample (BQS,
Schmidt \& Green 1983) consisting of $92$ QSOs with $B<16.4$. The model
consists of a double-power-law shape invariant with redshift,  with a steep
bright end $dN/dL\propto L^{-3.9}$, and a flatter faint end $dN/dL \propto
L^{-1.5}$. The evolution is simply represented by the so-called Pure Luminosity
Evolution (PLE) in which the characteristic luminosity ($L^\ast$) of the break
between the two power-laws evolves as a function of redshift according to
$L^\ast(z)\propto (1+z)^{3.4}$. 

In the last years a number of papers have been published suggesting that the
standard two power-law model in the framework of a pure luminosity evolution
may not be an adequate representation of the data (Goldschmidt {\it et al.}
1992; Miller {\it et al.} 1993; Hewett, Foltz \& Chaffee 1993; Hawkins \& Veron
1995). The most controversial points are the amount of flattening of the faint
part of the LF, first discovered by Boyle {\it et al.} (1988), and the shape of
the bright part which, in the last decade, has been based almost entirely on
the PG bright QSO survey data. 

In this paper we present a new determination of the QSO LF and its evolution.
We devote special care to the shape of the bright part of the LF, for which we
have derived new data from a bright QSO sample: the Homogeneous Bright QSO
Survey (HBQS, partly published in Cristiani {\it et al.} 1995). In the
computation we take also into account the most relevant already existing QSO
samples, in particular two new bright QSO surveys: the Edinburgh QSO Survey
(EQS, Goldschmidt {\it et al.} 1992), and the Large Bright QSO Survey (LBQS;
Hewett, Foltz \& Chaffee 1995 and references therein). 

\section{THE SAMPLES}

As discussed in section 4, we have used Monte Carlo simulations in order to
derive the best fit QSO LF. For this reason our analysis is mainly based on a
set of QSO samples for which the selection criteria and the photometric
properties are well defined or at least can be acceptably reproduced. In
addition to the knowledge of the K-correction and QSO colors, for a rigorous
determination of the LF we need the definition for each sample of: 
\begin{itemize}
\item the flux limits and selection criteria;
\item the distribution of the photometric errors as a function of apparent
magnitudes;
\item the galactic absorption;
\item the color equations of the photometric system.
\end{itemize}

The QSO LF will be computed in the Johnson B absolute magnitudes. In the
following color equations the ``natural'' photometric systems are marked with
primed letters, while the Johnson/Cousins system is represented with unprimed
letters. 

\subsection{The HBQS}

The Homogeneous Bright QSO Survey (HBQS) was started in 1989 in the framework
of an ESO Key-programme. The survey covers a total of 555 deg$^2$ subdivided in
22 ESO/Schmidt or UKST fields at high galactic latitude around the south
galactic pole ($b < -60^\circ$). 

Two Schmidt plates for each bandpass $U$ (IIa-O + UG1 or IIIa-J + UG1), $B'$
(IIa-O + GG385) or $B_J$ (IIIa-J + GG395), $V'$ (IIa-D + GG495), $R$ (IIIa-F +
RG630) or $OR$ (IIIa-F + OG590), and $I$ (IV-N + RG715) have been obtained at
the UKST or ESO/Schmidt telescope, usually within a few months interval, in
order to minimize the effects of variability. The plate material has been
scanned on the COSMOS microdensitometer. The resulting tables, containing the
instrumental magnitudes and other useful parameters for the objects detected in
each plate, have been merged together in one table per field. Only objects with
at least 4 detections (in 10 plates) have been included in this final table.
According to the ESO/UKSTU numeration, the 13 ESO fields which have been
selected in the $B'$ photometric system are: 290, 349, 355, 406, 407, 408, 410,
468, 469, 470, 474, 479, 534. The 9 UKST fields which have been selected in the
$B_J$ photometric system are: 287, 295, 296, 297, 351, 411, 413, SA94, SGP. 

We have selected as candidates all the UVx ``not-extremely-extended'' objects
satisfying a type of {\it modified Braccesi less-restricted} two-color
criterion (La Franca, Cristiani \& Barbieri 1992). The principal color for the
selection has been chosen to be the $U-B'$ or the $U-B_J$ according to the type
of the available plate material. The secondary color has been chosen to be
$B_J-R$ or $B_J-V'$ or $B'-V'$, always preferring the combination giving the
most reliable photometric accuracy. 

The adopted color equations (Blair \& Gilmore 1982) are, for the $B'$ (IIa-O +
GG385) system: 
$$  B = B' + 0.11(B-V); $$
while for the $B_J'$ (IIIa-J + GG395) system:
\begin{equation}
B = B_J' + 0.28(B-V)
\end{equation}

The magnitude intervals in which the selection is virtually complete vary from
field to field between $15.0<B<17.3$ and $15.0<B<18.8$. The sample extracted
from these data includes 327 QSOs ($M_B<-23$, $q_0 = 0.5$, and
$H_0=50~Km/s/Mpc$). Cristiani {\it et al.} (1995) describe in detail the
procedure followed in the data acquisition and reduction and report the first
results obtained in six fields ($\sim 150$ deg$^2$, the ones reaching fainter
limiting magnitudes), providing significant statistics in the range ($16.75 < B
< 18.75$). 

\subsection{The Edinburgh QSO Survey (EQS)} 

The Edinburgh bright UVx QSO Survey (EQS) has been carried out over 13 UKST
fields (the area is $333$ deg$^2$ large) with the $UB_JVRI$ multicolor system
(see Goldschmidt {\it et al.} 1992). The data reduction was very similar to
that of the HBQS. The photometric rms errors are of $0.09$ mag in each bandpass
in the magnitude interval $15<B<18$ were most of the QSOs are. The selection
criteria is fairly well represented by the equations: $$ U-B_J<-0.35 ~~~~ and
~~~~ B_J-R>0. $$ 

The intervals of magnitude for which a reliable sample can be extracted vary
from $15.0<B<17.5$ to $15.0<B<18.5$ in the different fields (Goldschmidt 1994).
For our computation of the LF we have used the bright ($B<16.5$) subsample of 8
QSOs with $z>0.3$ published by Goldschmidt {\it et al.} (1992). In this
subsample the AGN with $b=15.79$ and $z=0.380$ has been excluded because of a
wrong redshift identification (Goldschmidt 1994). 

\subsection{The LBQS} 

The Large Bright QSO Survey covers an effective area of $454$ deg$^2$ (Hewett,
Foltz and Chaffee 1995 and references therein). The photometry has been carried
out in the $B_J$ bandpass and the errors are on average of $0.10$ magnitudes,
taking also into account zero point errors and variations across UKST fields.
The sample is based on 1055 QSOs and AGNs, with apparent magnitudes in the
range $16.5<B_J<18.85$ and redshifts in the range $0.2<z<3.4$. The selection
has been carried using objective prism plates, looking for emission line and/or
blue continuum objects. Because of the selection of the blue continuum objects
the sample is complete in the redshift range $0.2<z<2.2$. It is not possible to
directly simulate the selection criteria of this survey, however in order to
check the effects in the LF by the inclusion of such a large database, we have
temptatively simulated the blue continuum selection as an UVx selection: $$
U-B_J< -0.30.$$ As for the EQS and HBQS samples the adopted color equation is
(1). In our computation of the LF we have restricted the LBQS to a subsample of
220 QSOs in the magnitude range $16.5<B<18.0$. 

\subsection{The SA94 Survey} 

The SA94 survey is an UVx (UBV) selected sample over an area of $10$ deg$^2$.
The QSOs have been selected according to the {\it less restricted} Braccesi UVx
criterion (La Franca {\it et al.} 1992). The completeness magnitude range is
$15.0<B<19.9$. The complete sample extracted from the data includes 97 QSOs.
The B magnitudes are directly published in the Johnson/Cousins system and no
color transformation is needed. 

\subsection{The Durham/AAT Survey} 

This sample comes from an UVx stellar object selection from UK Schmidt $U$ and
$B_J$ plates. The $B_J$ plates have been calibrated on the B magnitude system.
Therefore the magnitudes of this surveys are referred as $b$ magnitudes for 
which the color equation is (Boyle {\it et al.} 1990):
$$ B = b + 0.23(B-V - 0.9). $$

The whole catalogue containing 420 AGNs is subdivided in 34 circular fields
having a diameter of $40'$, corresponding to a total area of $11.9$ deg$^2$.
The completeness magnitude ranges vary from field to field, the faintest limit
magnitude reaching $b=20.9$. Different UVx selection limits have also been
adopted in the various fields (see Boyle {\it et al.} 1990, Table 2). On
average they correspond to $ u-b<-0.40.$ In the redshift interval $0.3<z<2.2$
the sample consists of 353 QSOs. 

\subsection{The ESO/AAT} 

This is a faint $UJRI$ survey of 66 AGNs in the fields 855, 861, 864 (Boyle,
Jones and Shanks 1991), where $J'$ is a IIIa-J + GG385 system. The color
equation is (Kron 1980): $$  B = J' + 0.23(B-V). $$ The survey  covers an
effective area of $0.85$ deg$^2$ from $J=18.0$ down to $J=21.75-22.0$, and the
completeness redshift range is $0.6<z<2.9$. The typical rms scatter in the
calibration is 0.09 mag for the J bandpass, 0.15 mag for R passband, and 0.12
mag in the U bandpass. The candidates were selected among all the stellar
objects lying blueward of the stellar locus in either $U-J$ or $J-R$ diagrams.
At $z<2.2$ the subsample consists of 41 QSOs whose selection is well
represented with $$ U-J<-0.30.$$ 

\subsection{The (ZM)$^2$B Survey} 

This is a faint UVX survey of 54 AGNs from UJF plates taken at the ESO 3.6m
telescope (Zitelli {\it et al.} 1992). The survey cover an area of 0.69 deg$^2$
down to $J=20.85$ and 0.35 deg$^2$ down to $J=22.0$. In the redshift range
$0.3<z<2.2$ the complete sample consists of 19 QSOs. The selection criterion is
well represented by $$ U-J<-0.30.$$ As in the previous survey the color
equation is: $$  B = J' + 0.23(B-V). $$

\section{THE QSO COUNTS}

\begin{figure}[t]
\epsfxsize=88truemm
\epsffile{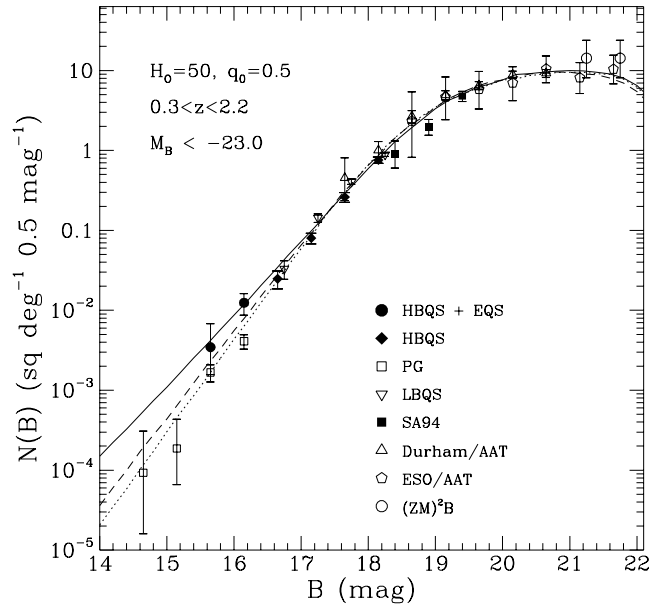}
\caption{The QSO number counts for $0.3<z<2.2$. The continuous
line represents 
the best fit LDLE model (model C in Table 2), the dashed line represents our
best fit PLE model (model B in Table 2), while the dotted line represents the
best fit PLE model (model A in Table 2) by Boyle {\it et al.} (1992).
The magnitudes
have all been corrected for galactic extinction according to Burstein and
Heiles (1982).} 
\end{figure}

The B number-magnitude counts for all the samples used in the computation of
the optical LF are shown in Fig. 1. In the following we will restrict our
analysis to $0.3<z<2.2$ because this is the only interval in which the
selection criteria allow a reliable and homogeneous processing of the data. The
magnitudes have all been corrected for galactic extinction according to
Burstein \& Heiles (1982). The QSO surface densities from the HBQS and EQS are
also shown in Table 1. As first indicated by Goldschmidt {\it et al.} (1992),
for magnitudes brighter than $B=16.4$ the QSO surface density turns out to be a
factor 2.5 higher than what measured by the PG survey (Schmidt \& Green 1983).
The data from the HBQS collect 7 QSOs at magnitudes brighter than $B =16.4$,
corresponding to a surface density of $0.013^{+.007}_{-.005}~deg^{-2}$. With
the addition of the EQS, the surface density at $B<16.4$ becomes $0.016 \pm
0.005 ~ deg^{-2}$, corresponding to a total of 14 QSOs over an area of $888~
deg^{2}$. The PG sample lists 44 QSOs at $B<16.4$. In this range the surveyed
area changes rapidly as a function of the limiting magnitude. Combining the
number of discovered QSOs with the areas according to Table 1 of Green, Schmidt
\& Liebert (1986), a QSO surface density of $0.006 \pm 0.001~deg^{-1}$ is
obtained, a factor of 2.7 lower (significant at 2$\sigma$ level) than the
EQS+HBQS estimate. The presence of incompleteness in the PG survey has been
investigated by Wampler and Ponz (1985). It probably originates for QSOs with
$U-B$ colors near the selection limit because of large rms errors. This effect
would lead to incompleteness that varies with redshift, with the consequence
that, in particular in the range $0.4<z<0.7$, it is systematically
underpopulated at a given magnitude. This hypothesis appears corroborated by
the fact that the PG survey, in the area covered to a limit $B = 16.4$, has
discovered 46 QSOs with $z<0.3$, in addition to the 44 with $z>0.3$ that are
relevant for the comparison with the current sample. For these reasons we have
excluded the PG sample from our subsequent analysis of the evolution of the
LF. At fainter magnitudes, in the interval $16.5<B<18.5$, the HBQS shows a
fairly good agreement with the counts of the LBQS.

\placetable{tbl-1}
\begin{table}
\dummytable\label{tbl-1}
\end{table}

\section{THE LUMINOSITY FUNCTION} 

\subsection{Deterministic Derivations} 

The QSO LF $\Phi(M,z)$ is the measure of the number of QSOs per unit of
commoving volume with magnitudes in the interval ($M, M+dM$) as a function of
redshift $z$. In the redshift interval $0.3<z<2.2$ the general behaviour of the
QSO LF has been usually parameterized with a double power-law Pure Luminosity
Evolution (PLE) model $ \Phi (M_B,z) =$ $$ { {\Phi^\ast} \over {
10^{0.4[M_B-M_B(z)](\alpha + 1)} + 10^{0.4[M_B-M_B(z)](\beta + 1)} } } $$ where
$\alpha$ and $\beta$ correspond to the faint-end and bright-end slopes of the
optical LF, respectively. With this parameterization the evolution of the LF is
uniquely specified by the redshift dependence of the  break magnitude, 
$$ M_B (z) = M_B^{\ast} - 2.5k\log(1+z),$$
\noindent
corresponding to a power-law evolution in the optical luminosity,
$L^{\ast}\propto (1+z)^k$. $\Phi (M_B,z)$ is expressed in units of $Mpc^{-3}
mag^{-1}$ (see Boyle {\it et al.} 1988). 

The empirical way of deriving the LF is by collecting all the available QSO
samples, computing the transformations to a common photometric bandpass, and
measuring the QSO space densities in $\Delta z$, $\Delta M$ bins using the
$1/V_a$ estimator of the "coherent" method of Avni \& Bahcall (1980); where
$V_a$ is the {\it accessible volume} computed for every object over all the
samples and not just the sample in which the object originated. This method has
a number of disadvantages as it is affected by: 

\begin{enumerate} 

\item 
the uncertainties due to the binning of the data. As the number of QSOs
in the available complete samples is still small, it is necessary to use
relatively large bin sizes on the $M_B-z$ plane within which the evolutionary
effects are not negligible. In this way the non-linear dependencies of the QSO
LF on $M_B$ and $z$ cause spurious trends on the derived shape of the LF. 

\item 
the influence of the photometric errors. Cavaliere, Giallongo and
Vagnetti (1989) pointed out how the uncertainties introduced in the statistical
samples of QSOs by photometric variances and from the fluctuation of the sky
variations cause an overestimation of the evolution timescales (see also
Giallongo and Vagnetti 1992, and Francis 1993). 

\item 
the spread in the assumed mean spectral index of the QSOs used to compute
the K-correction. If the contribution of the emission lines to the broadband
flux can be neglected, the formula relating the absolute and apparent blue
magnitudes is (Schmidt \& Green 1983) 
$ M_B =$ $$ B - 5 \log [A(z)c/H_0]+2.5(1+\gamma)\log(1+z)-25,$$ 
where $A(z)c/H_0$ is the bolometric luminosity distance, and the extrapolation
from the observed to the rest-frame B band assumes a power-law representation
of the QSO continuum $f_{\nu}\propto \nu^{- \gamma}$. When quantifying the
evolution of the population between two redshifts, if the extrapolation is
large, it is important to know the spread of the distribution of $\gamma$. Any
distribution of the mean spectral index is reflected in an uncertainty in the
absolute magnitudes, which increases with redshift and introduces an artificial
luminosity evolution of the population. The dispersion in the power-law index
$\gamma$ in the rest-frame ultraviolet is substantial (Sargent, Steidel and
Boksenberg 1989; Francis {\it et al.} 1991; Schneider {\it et al.} 1991), with
estimates of $\sigma_{\gamma}=0.3-0.6$. For a sample of QSOs measured at the
same observed wavelength the differential effect between two redshifts $z_1$,
$z_2$ is $$\Delta M_B = 2.5\Delta \gamma \log [(1+z_2)/(1+z_1)].$$ In the
redshift range of our analysis $0.3<z<2.2$, changing from $z=0.3$ to $z=2.2$
results in a uncertainty $\sigma_{M_B} \simeq \sigma_{\gamma}$. A value $\Delta
\gamma = 0.5$ would correspond to an error in the space density of a factor 3
in the bright part of the LF assuming a power-law index $\beta=-3.5$. In the
case of two QSOs of the same redshift and apparent magnitude $B$ but different
spectral index, the difference in absolute magnitude is $$\Delta M_B =
2.5\Delta \gamma \log (1+z).$$ Changing from $\gamma=0$ to $\gamma=1.5$ would
produce at $z=0.4$ a difference in absolute magnitude of $\Delta M_B=0.4$ mag.
At $z=2.2$ the difference becomes $1.9$ mag corresponding to an error in the
space density of a factor 100 in the bright part of the LF (assuming a power
law index $\beta=-3.5$). The final result of this effect is a blurring of the
LF. 

\end{enumerate}

The ``standard'' maximum likelihood techniques (Marshall {\it et al.} 1983,
Boyle, Shanks and Peterson 1988) do not need binning of the data and
overcome the disadvantage described in item (1), but do not solve problems (2)
and (3). 

In order to take into account also these latter two effects, it has been
proposed to compute a detection probability as a function of apparent
magnitude, redshift and Spectral Energy Distribution ($SED$) (see Warren,
Hewett \& Osmer 1994). The detection probability $p(m,z,SED)$ is derived
computing the fraction of QSOs that would have been selected by each survey
according to the selection criteria, photometric errors, and the spectral
index. Thus, for each redshift and magnitude interval, the space density is
evaluated as $\Sigma 1/\int p(z)dV_a$. Eventually, in order to determine the
range of parametric luminosity-function models that are consistent to the data,
the observed and expected number of QSOs are compared through maximum
likelihood or least $\chi^2$ techniques. 

\subsection{The Monte Carlo Simulations}

The above described procedures do not lead to a fully satisfactory and
rigorous determination of the LF. The photometric errors and the spread in the
QSO SED properties make the nature of the problem intrinsically stochastic.
Even taking into account the influence of different SEDs in the computation of
the absolute luminosity, the problem remains basically unsolved. For example,
because of the effects of the photometric errors, two QSOs observed at the same
apparent magnitude and with the same SED and redshift may be produced by
objects of different absolute magnitudes. This means that it is impossible to
establish a bi-unique (one-to-one) relationship between the observed apparent
magnitudes and the absolute magnitudes as a function of $z$ and $SED$. 

For these reasons we have followed a different approach, simulating, with
Monte Carlo techniques, what happens when a selection of a sample is
made. For each sample the following procedure has been applied: 

\begin{enumerate}

\item
Assuming a theoretical LF distribution $\Phi(M_B,z)$, a number (proportional to
the surface area of the survey) of QSOs (i.e. pairs $M_B, z$) have been
randomly extracted; 

\item
The apparent magnitudes of the QSOs have been computed following the
K-correction according to Cristiani \& Vio (1990), and deriving the average
galactic absorption $A_B$ from the HI maps of Burstein and Heiles (1982). The
resulting apparent B magnitude is: 
$$ B = M_B + 5 \log [A(z)c/H_0] + 25 + K(z) + A_B + \epsilon(B)$$
where $\epsilon(B)$ simulates a noise with Gaussian distribution with variance
$$ \sigma^2_B = \sigma^2_{phot}(B) + \sigma^2_{B,\gamma}.$$ $ \sigma^2_B$ takes
into account both the photometric errors of the survey $\sigma_{phot}(B)$ and
the apparent magnitude dispersion due to the QSO spectral slope dispersion
$\sigma_{\gamma}$ 
$$\sigma_{B,\gamma} = 2.5\sigma_{\gamma} \log (1+z).$$

\item
The apparent colors have been generated following the average QSO
colors-redshift dependence $F_{1,2}(z)$ (Cristiani \& Vio 1990). Between each
couple of bandpasses centered at $\lambda_1$ and $\lambda_2$ the resulting
apparent color is: 
$$ C_{1,2} = F_{1,2}(z) + \epsilon_{F_{1,2}}$$
where $\epsilon_{F_{1,2}}$ simulates a noise with Gaussian distribution with
variance 
$$ \sigma^2_{F_{1,2}} = \sigma^2_1 + \sigma^2_2 + \sigma^2_{F,\gamma}.$$ 
$ \sigma^2_{F_{1,2}}$ takes into account the photometric errors of the two
bandpass $\sigma_1$ and $\sigma_2$ respectively, and the QSO color dispersion
due to the spread $\sigma_{\gamma}$ in the slope of the spectra 
$$\sigma_{F,\gamma} = 2.5\sigma_{\gamma} \log {\lambda_2 \over \lambda_1}.$$

Note that the error on the color $\sigma_{F,\gamma}$ is correlated with the
error on the apparent magnitude $\sigma_{B,\gamma}$. For each simulated QSO 
the same value of $\gamma$ has been used to compute the apparent magnitude and
colors. 

\item
The flux limits and selection criteria of the survey have been applied to the
resulting magnitudes and colors in order to select the ``observed'' QSOs. 

\end{enumerate}
In addition to statistical errors, the presence of systematic errors in the QSO
photometry could induce biases in the Monte Carlo simulations. In particular
the applicability of the Blair and Gilmore (1982) transformations, which have
been calibrated on a range of colors ($-0.4\leq U-B\leq 2.0$), not entirely
relevant for low-redshift QSOs, is not straightforward. It is reassuring that
the object for which the effect might be significant for the
selection/non-selection, are those closer to the stellar locus, where the color
transformations are still reliable. 

The observed QSO magnitudes and colors are also affected by variability that in
principle gives origin to at least two effects:

\begin{enumerate}

\item
The shape of the LF is modified (blurred) by variability. We will derive the
QSO LF as it is observed at a given instant, so we will neglect the effect from
this point of view. How such a ``snapshot'' LF is translated into an
``average'' LF, describing the distribution of the average luminosities of QSOs
on long timescales, is matter for future variability studies. 

\item
Variability is going to affect the observed colors of QSOs, acting as an
additional photometric error, if the photometric material in the different
colours is obtained at different times. The critical timescale above which the
effects of variability become non-negligible can be set at about one year in
the observer's rest frame (Cristiani {\it et al.} 1996). However, the effects
of variability can be modeled with a spectral slope variable in time. In
particular this  can reproduce the observed  dependence of the variability on
the wavelength (Di Clemente {\it et al.} 1996, Cristiani {\it et al.} 1996). In
this a variable spectral slope could manifest itself in an ensemble of QSOs as
a spread in the observed spectral index, that is already included in our
computations. Therefore we have not introduced further effects of the
variability on the spread of the QSO colors. 

\end{enumerate}

\subsection{Results}

As discussed in section 2, in order to compute the best fit LF we have added
the QSOs of the HBQS and EQS to all the optically selected samples for which it
was possible a reliable assessment of the selection criteria and the
photometric quality. In this way, a total sample of 1022 QSOs has been
obtained. The fits have been carried out by minimizing the $\chi^2$ statistics
derived from the comparison of the observed $(B, z)$ distribution with 2000
simulations of each theoretical LF model. Table 2 summarizes the results for a
number of different models. 

The errors quoted for the parameters are 68 per cent ($1~\sigma$) confidence
intervals. They correspond to variations of $\Delta \chi^2 = 1.0$, obtained
perturbating each parameter in turn with respect to its best-fit value, and
looking for a minimization with the remaining parameters free to float. 

The significance of the fitting has also been tested for goodness-of-fit in the
lowest redshift interval $0.3<z<0.6$, in which the fitting probabilities have
been computed using the two-dimensional Kolmogorov-Smirnov (2D KS) test by
Fasano and Franceschini (1988). See Press {\it et al.} (1992) for a complete
description of the algorithm.

\placetable{tbl-2}
\begin{table}
\dummytable\label{tbl-2}
\end{table}

In the following we first focus our discussion on a $q_o =0.5$ universe, we
will later see how our results change for $q_o = 0.1$. 

In Fig. 2 our best fit PLE model (B) is shown. The data points, representing
the estimates of the QSO space densities, are compared to the fitted LF. The
observed QSO space densities have not been directly computed through the
$1/V_a$ technique, that, due to the binning of the data, can give origin to
spurious trends on the shape of the LF (see sub-section 4.1). For this reason
we have preferred to exploit our Monte Carlo simulation code in order to {\it
visualize} the LF. The value of each observational point in Fig. 2 has been
computed by comparing, in each absolute magnitude/redshift bin, the number of
observed QSOs, $N_o$, with the number of QSOs simulated with the fitting LF,
$N_s$. Let us call $\Phi_o$ and $\Phi_s$ the corresponding luminosity functions
and $\Psi(M,z)$ the selection function. In each bin we have: 
$$ {{N_o}\over{N_s}} = {{\int \Psi(M,z) \Phi_o(M,z) dMdz}\over{\int
\Psi(M,z) \Phi_s(M,z) dMdz}}, $$
where the integrals are computed over the corresponding bin. In our case, {\it
in each bin}, we can assume that the shape of the observed and fitting LFs are
very similar, apart from a constant factor $k$, which varies from bin to bin:
$\Phi_s \simeq k \Phi_o$. Using this assumption we can write: 
$${{N_o}\over{N_s}} \simeq {{\Phi_o(M,z)}\over{\Phi_s(M,z)}}, $$
for all values $M$ and $z$ in the bin. Consequently, for proper visualization
of the data, the observed QSO densities have been computed by multiplying the
corresponding value of the simulating LF, $\Phi_s$, at the central magnitude
and redshift of the bin, by the ratio between the number of observed and
simulated QSOs in the corresponding bin: 
$$ \Phi_o(M,z) \simeq{{N_o}\over{N_s}}\Phi_s(M,z). $$
The 1 $\sigma$ upper and lower limits have been estimated with the Poissonian
statistics uncertainties on $N_o$. As described above, the LF fitting technique
is completely independent from these assumptions, being based on a $\chi^2$
minimization of the difference between $N_o$ and $N_s$ in bins of $B$ and $z$. 

\begin{figure}[t]
\epsfxsize=88truemm
\epsffile{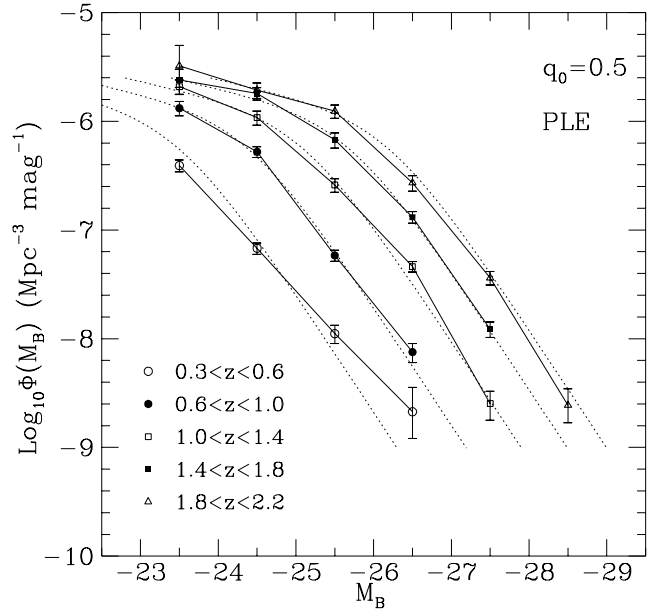}
\caption{The QSO luminosity function. The points connected with a continuous
line represent the observations (see text for details about their
derivation). The dashed lines correspond to the best-fitting
Pure Luminosity Evolution
(model B in Table 2). Error bars are based on Poisson statistic and
correspond to 68 per cent ($1 \sigma$) confidence intervals.}
\end{figure}

Our PLE model B gives a satisfactory fit of the ($B, z$) distribution with a
global $\chi^2$ probability of 0.21. The PLE model A by Boyle (1992), compared
with the present data, provides a lower but still acceptable probability of
0.12, however with one additional parameter, $z_{cut}=1.9$, representing the
redshift at which the luminosity evolution ``switches off''. At redshift
greater than $z_{cut}$ no further luminosity evolution takes place. Our B model
has a flatter bright slope $\beta$ ($-3.7$ compared to $-3.9$) than the A
model. The evolution parameter $k$ is also smaller ($3.26$ compared to $3.45$).

This difference in the parameter $\beta$ is originated by a larger density of
luminous QSOs (expecially at $z<0.6$) in comparison with the previous data
derived from the PG sample (compare Fig. 2 with Fig. 3b of Boyle 1992). The
higher evolutionary rate $k$ determined by Boyle (1992) is a result of the
introduction of the redshift cut off at $z_{cut}=1.9$, which with our data
results unnecessary. 

Both A and B models provide an inadequate simulation of the distributions of
the observed data for the low redshift domain, $0.3<z<0.6$. For this subsample
the 2D KS statistics test rejects model A at the 0.01 level and model B at 0.02
level. At magnitudes brighter than $M_B=-25$, in the interval $0.3<z<0.6$, 32
QSOs are observed, while 16 and 19 QSOs are predicted by model A (a $4~\sigma$
discrepancy) and B (a $3~\sigma$ discrepancy) respectively. But as the low
redshift ($z<0.6$) subsample contain only 5 per cent of the complete data set,
in a global comparison of the whole data sample in the interval $0.3<z<2.2$,
the models follow the evolution of the larger fraction of QSOs at higher
redshift, allowing the $\chi^2$ probability of the global fit to reach
satisfactory levels. 

No significant difference is obtained by fitting the data excluding each of the
three bright samples (HBQS, EQS and LBQS) in turn (models D, E and F). 

It is interesting to see how the inclusion of a spread in the theoretical
average QSO spectral slope modifies the best fit luminosity function. As
expected (Giallongo and Vagnetti 1992), larger values of the spread of the slope
correspond to slower luminosity evolution parameters and steeper luminosity
functions (model G with $\sigma_\gamma = 0.3$, and model H with $\sigma_\gamma
= 0.5$). 

In order to fully represent the data in the redshift range $0.3<z<2.2$ we tried
several functional modifications of the PLE. As first step we tried, without
significant results, to add one or at maximum two parameters in order to take
into account of a dependence of the bright slope $\beta$ with redshift. 

The best description of the observed data has been obtained by decreasing at
low redshift the luminosity evolution of the bright QSOs. This has been
obtained by including a dependence on the redshift and absolute magnitude of
the evolution parameter $k$ such as: 
\begin{eqnarray}
\label{eq:LDLE}
for~M_B \leq M^{\ast}&:&~ k = k_1 + k_2 (M_B-M^{\ast})e^{-z/{.40}} \\
\nonumber for~M_B > M^{\ast}&:&~ k = k_1 
\end{eqnarray}
where $M^{\ast}$ is the magnitude of the break in the two power-law shape of
the LF. This luminosity dependent luminosity evolution (LDLE) model (model C in
Table 1) has resulted in a better fit of the data (see Fig. 3) giving a
$\chi^2$ test probability of 0.49 in the whole $B, z$ plane, and an acceptable
2D KS probability of 0.09 in the redshift interval $0.3<z<0.6$. With this
model, in this redshift  interval and for magnitudes brighter than $M_B=-25$,
29 QSOs are expected in comparison with the 32 observed. 

\begin{figure}[t]
\epsfxsize=88truemm
\epsffile{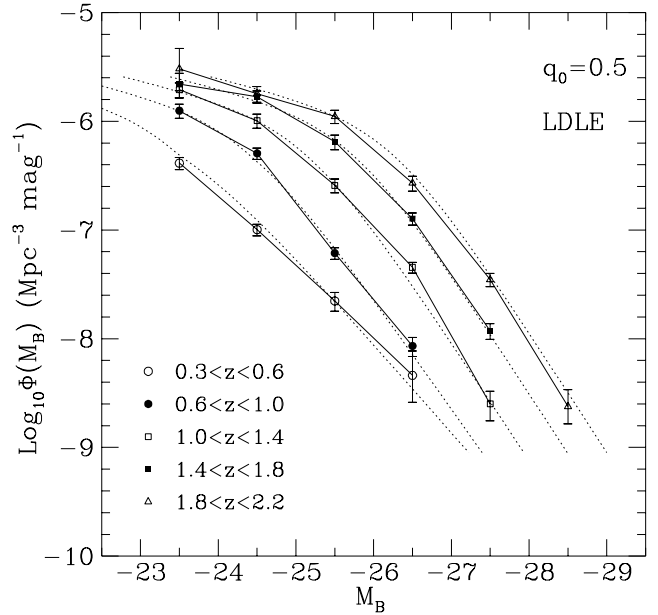}
\caption{
The QSO luminosity function. The points connected with a continuous
line represent the observations (see text for details about their
derivation). The dashed lines correspond to the best-fitting
Luminosity Dependent Luminosity Evolution (model C in Table 2).
Error bars as in Fig. 2.} 
\end{figure}

In this way, as shown in Fig. 1, the LDLE reproduce the higher counts of bright
QSOs at $B<16.4$, discovered by the HBQS and EQS surveys, much better than the
PLE models: a surface density of $0.017~deg^{-2}$ is predicted with respect to
$0.016~deg^{-2}$ observed. The A model by Boyle (1992) predicts a surface
density of $0.008~deg^{-2}$. 

It is important to notice that both the PLE and the LDLE models do not
reproduce a 2$\sigma$ significant observed underdensity of faint QSOs
($-23>M_B>-25$) in the redshift range $0.3<z<0.6$. In this bin, with the PLE B
model, 125 QSOs are observed and 161 expected. It is possible to find less
significant indications of this feature in the data analyzed by Boyle (1992).
We examined the possibility that this observed underdensity of the faint part
of the luminosity function at low redshift is due to an underestimated
incompleteness of the UVx optical selection method at low redshift. For this
reason we have carried out simulations assuming a redder $U-B$ QSO color of
$0.2$ magnitudes in the interval $0.3<z<0.8$. The effects were still
insufficient to account for the underdensity at low redshift (125 QSOs observed
and 154 expected). As our best fit probabilities are satisfactory, we preferred
not to model the low redshift faint feature. It is worth to notice that the
determination of the local ($z=0$) AGN (Seyfert 1 plus QSOs) LF from the
Hamburgh Bright QSO Survey (Wisotzki {\it et al.} 1996a) shows similar trends,
with evidences of higher densities at bright magnitudes and lower densities at
faint magnitudes in comparison with the de-evolved PLE LF by Boyle (1992)
(Wisotzki {\it et al.} 1996b; K\"ohler 1996). 

Adopting a $q_o = 0.1$ Universe, as shown in Table 2, the fitting of the data
with the PLE model (I) by Boyle (1992) is rejected at 0.004 confidence level.
Our PLE model (L) obtains an acceptable global representation of the data
($\chi^2$ probability 0.36), slightly better than the PLE fit in a $q_o = 0.5$
Universe. However, in the $0.3<z<0.6$ redshift range the PLE still
underestimates the expected number of QSOs, although the discrepancy is less
serious than in the $q_o = 0.5$ case: 44 QSOs with $M_B<-25$ are observed
versus 33 predicted in our L model (24 in the I model). The introduction of a
LDLE parameterization according to Eq.~\ref{eq:LDLE} provides satisfactory
representation of the observations in the low-$z$ domain also for $q_o = 0.1$. 

\section{DISCUSSION}

What is the origin of the flattening of the luminosity function at
low-redshift and is it confirmed by observations in other wavebands? 

In the last years evidence has been produced that optically selected QSOs have
a bimodal distribution of radio flux densities, and that the fraction of radio
loud QSOs decreases with increasing redshift and decreasing absolute optical
luminosity (Miller {\it et al.} 1990; Visnovsky {\it et al.} 1992; Schneider
{\it et al.} 1992; Padovani 1993; La Franca {\it et al.} 1994). 

La Franca {\it et al.} (1994) have determined that the radio-loud fraction is
substantially higher for QSOs with $z<1$, a result essentially due to the
higher overall radio-loud fraction in the predominantly low redshift PG sample.
The radio-loud fraction has also been found to be larger at bright absolute
magnitudes, regardless of whether or not the PG sample is included in the
analysis. La Franca {\it et al.} (1994) have been able to reproduce the data
with a radio-loud optical luminosity function (OLF) characterized by a lower
space density, a similar shape and a smaller evolutionary parameter $k$
($2.7<k<3.1$ ) with respect to the parent {\it total} (radio-quiet plus
radio-loud)
QSO population. According to this picture the flattening of the bright
part of the OLF corresponds to an increase of the radio-loud OLF contribution. 

However, the decrement of the radio-loud fraction with redshift relies
essentially on the PG sample and has not been confirmed by a study of 256 QSOs
in the LBQS (Hooper {\it et al.} 1995). Nevertheless there is a very small
overlap in the magnitude redshift plane between the LBQS and the PG sample: the
LBQS QSOs are too faint and too high-redshift to be compared with the PGs. Our
observed overdensity of low redshift ($0.3<z<0.6$) QSOs involves the QSOs with
$B<16.5$ and $M_B<-25$ but in the LBQS there are only two such QSOs with
$B<16.5$, to be compared with the 45 QSOs of the PG sample. In the EQS plus
HBQS combined sample there are 16 QSO brighter than 16.5. It becomes crucial to
obtain radio data of this QSO subsample. 

Other information on the evolution of the LF and on the dependence of the radio
loud fraction on magnitude and redshift can be derived from QSO samples
selected in the soft X-rays. 

A study of the EMSS AGN sample by Maccacaro {\it et al.} (1991) and Della Ceca
{\it et al.} (1992) has shown that a PLE model with a power-law dependence on
the redshift, $L_x(z)=L_x(0)(1+z)^k$, provides the simplest description of
data. Their best-fitting evolution rate ($k_x=2.56$) is almost one unit lower
than the optical best fit. Boyle {\it et al.} (1993) used a new sample of faint
AGNs selected by ROSAT, combined with the EMSS sample, to derive a value of
$k=2.8$. Using the same data Franceschini {\it et al.} (1994) showed that the
evolution rate of X-ray selected AGNs could have been underestimated and
actually be comparable to that of the QSOs in the optical, with a linear
scaling of the emission in the two bands. Boyle {\it et al.} (1994) using an
enlarged sample of 107 faint AGNs selected by ROSAT, combined with EMSS sample,
derived an evolution $L_x(z)=L_x(0)(1+z)^{3.25}$ at $z<1.6$, quite similar to
that of optically selected QSOs. The latter two analyses favour a picture in
which the same QSO population is observed, both in the soft-X and in the
optical. If this is true, at low $z$, bright X-ray selected samples should
show a flattening of the bright part of the OLF and an increase of the radio
loud fraction. Unfortunately the brightest X-ray selected QSO sample, the EMSS,
is not bright (large) enough to probe this domain of the OLF. In the interval
$0.1765<z<0.4286$ the brightest significant bin is at $LogL_X = 45~erg s^{-1}~
(0.3-3.5~KeV)$ (Maccacaro {\it et al.} 1991), which corresponds to about
$M_B=-25$ ($q_0=0.1$, see Fig. 5 in Franceschini {\it et al.} 1994, and
La Franca {\it et al.} 1995), in the interval $0.4286<z<08182$ the EMSS reaches
$LogL_X = 45.5~erg s^{-1}~ (0.3-3.5~KeV)$, corresponding to about $M_B=-26$.
It is interesting to notice that Jones {\it et al.} (1996), at bright
magnitudes in the interval $0.0<z<0.4$, show a flatter X-ray LF with
respect to their PLE model (see their Figure 5). 

Della Ceca {\it et al.} (1994) have analyzed the radio properties of the EMSS:
they find that in a subsample of optically luminous QSOs, 30\% of the objects
are radio-loud, while in the complementary faint subsample the fraction of
radio-loud drops to 3\%. No evidences are found that the evolutionary
properties of the radio-loud objects differ significantly from the radio-quiet
ones. However, the shape of the de-evolved XLF of the two classes appears to be
different and a flattening of the XLF of the radio-loud subsample is visible
for $L_X(z=0)<3\times 10^{44}~erg s^{-1}$. 

Ciliegi {\it et al.} (1995) have obtained VLA observation of a ROSAT
sample of 80 faint AGNs. They combined their sample with EMSS data and found
that, in the case of a PLE, $L_x(z)=L_x(0)(1+z)^k$, $k=2.43\pm 0.26$ and
$k=2.71\pm 0.10$ for the radio-loud and radio-quiet population respectively.
Although not significant, the difference hints a slower evolution of the radio
loud population.

In summary, at the moment data from different wavebands suggest that: 

\begin{enumerate}

\item There is a flattening of the bright part of the OLF at low redshift;

\item There is an increase of the radio-loud fraction at $M_B < -25$ both in
X-ray and optically selected samples; 

\item The apparent luminosity evolution of the radio-loud population is
slower both in optically and in X-ray selected samples; 

\item The same QSO population is selected in the optical and soft-X domains,
and X-ray selected QSOs follow a PLE evolution similar or slower than that of
the optically selected QSOs; 

\item The X-ray selected samples do not have a sufficient coverage of the $L,
z$ plane to allow a meaningful cross-check of the flattening in the bright part
of the low-{$z$} OLF. 

\end{enumerate}

In this way the observed overdensity of low redshift bright QSOs and the
related flattening of the LF could be originated by the contribution of a
slower evolving population of radio-loud QSOs. To put this scenario on a firmer
basis it is necessary to obtain radio data for: 
1) a new bright, wide field ($\sim 10000~deg^2$) QSO sample, in the interval
$13 < B <16$, and 
2) a wide field bright X-ray selected QSO sample. 
The former will be provided by the Sloan Survey (Gunn and Weinberg 1995), while
the Rosat All Sky Survey seems to be ideally tailored for the latter. 

\section{acknowledgements}

FLF acknowledges financial support from CNR, the hospitality of the {\it
Dipartimento di Astronomia dell'Universit\`a di Padova}, and of the {\it
Istituto di Astronomia dell'Universit\`a "La Sapienza" di Roma}. 
This work was partially supported by the ASI contracts 92-RS-102, 94-RS-107 and
95-RS-38 and by the HCM programme of the European Community.

\vskip 2 truecm

{\sc TABLE} 1. QSO counts and surface densities.

{\sc TABLE} 2. ``Best fit'' parameters for luminosity function models.


\begin{thebibliography}{}

   \bibitem{}
Andreani S., Cristiani S. 1992, ApJ, 298, L13

   \bibitem{}
Avni Y., Bahcall J.N. 1980, ApJ, 235, 694

   \bibitem{}
Blair M., Gilmore G. 1982, PASP, 94, 742

   \bibitem{}
Boyle B.J., Shanks T., Peterson B.A. 1988, MNRAS, 235, 935 

   \bibitem{}
Boyle B.J., Fong R., Shanks T., Peterson B.A. 1990, MNRAS, 243, 1 

   \bibitem{}
Boyle B.J., Jones L.R., Shanks T. 1991, MNRAS, 251, 482 

   \bibitem{}
Boyle B.J. 1992, in "Texas/ESO-CERN Symposium on Relativistic Astrophysics,
Cosmology and Particle Physics", ed(s) Barrow J.D., Mestel L. and
Thomas P.,  Ann. N.Y. Acad. of Sci., 647, 14

   \bibitem{}
Boyle B.J., Griffiths R.E., Shanks T., Stewart G.C., Georgantopoulos I. 1993,
MNRAS, 260, 49 

   \bibitem{}
Boyle B.J., Shanks T., Georgantopoulos I., Stewart G.C., Griffiths R.E. 1994,
MNRAS, 271, 639 

   \bibitem{}
Burstein D., Heiles C. 1982, AJ, 87, 1165

   \bibitem{} 
Cavaliere A., Giallongo E., Vagnetti F. 1989, AJ, 97, 336

   \bibitem{} 
Ciliegi P., Elvis M., Wilkes B.J., Boyle B.J., McMahon R.G., Maccacaro T.
1996, MNRAS, 277, 1463 

   \bibitem{}
Comastri A., Setti G., Zamorani G., Hasinger G. 1995, A\&A, 296, 1

   \bibitem{} 
Cristiani S., Vio R. 1990, A\&A, 227, 385

   \bibitem{} 
Cristiani S., La Franca F., Andreani, P., {\it et al.} 1995, A\&AS, 112, 347

   \bibitem{} 
Cristiani S., Trentini S., La Franca F., Aretxaga I., Andreani P., Vio R.,
Gemmo A. 1996, A\&A, 306, 395

   \bibitem{} 
Della Ceca R., Maccacaro T., Gioia I.M., Wolter A., 
Stocke J.T. 1992, ApJ, 389, 491

   \bibitem{} 
Della Ceca R., Zamorani G., Maccacaro T., Wolter A., Griffiths R.,
Stocke J.T., Setti G. 1994, ApJ, 430, 533

   \bibitem{} 
Di Clemente A., Giallongo E., Natali G., Trevese D., Vagnetti F. 1996,
ApJ, 463, 466

   \bibitem{} 
Efstathiou G., Rees M.J. 1988, MNRAS, 230, 5P

   \bibitem{} 
Fasano G., Franceschini A. 1988, MNRAS, 225, 155

  \bibitem{} 
Franceschini, A., La Franca, F., Cristiani, S., Mirones, J.M. 1994,
MNRAS, 269, 683

   \bibitem{} 
Francis P.J., Hewett P.C., Foltz C.B., Chaffee F.H., Weymann R.J., and
Morris S.L. 1991, ApJ, 373, 465

   \bibitem{} 
Francis P.J. 1993, ApJ, 407, 519
 
   \bibitem{} 
Giallongo E., Vagnetti F. 1992, ApJ, 396, 411

   \bibitem{} 
Goldschmidt P., Miller L., La Franca F., Cristiani S. 1992, MNRAS, 256, 65p 
 
   \bibitem{}
Goldschmidt P. 1994, PhD Thesis, Department of Astronomy, University of
Edinburgh

   \bibitem{}
Green R.F., Schmidt M., Liebert J. 1986, ApJS, 61, 305

   \bibitem{Gunn and Weinberg 1995}
Gunn J., Weinberg D. 1995, in Wide Field Spectroscopy and the Distant
Universe, Proc. of 35th Herstmonceux Conference, ed. Maddox and
Arag\'on-Salamanca (World Scientific, Singapore), p. 3

   \bibitem{}
Haardt F., Madau P. 1996, ApJ, 461, 20

   \bibitem{} 
Haehnelt M.G., Rees M.J. 1993, MNRAS, 263, 168

   \bibitem{} 
Hartwick F.D.A., Schade D. 1990, ARA\&A, 28, 437   
 
   \bibitem{} 
Hawkins M.R.S., V\'eron P. 1995, MNRAS, 275, 1102

   \bibitem{} 
Hewett P.C., Foltz C.B., Chaffee F.H. 1993, ApJ,  406, L43 

   \bibitem{} 
Hewett P.C., Foltz C.B. 1994, PASP,  106, 113 

   \bibitem{} 
Hewett P.C., Foltz C.B., Chaffee F.H. 1995, AJ, 109, 1498 
 
   \bibitem{}
Hooper E.J., Impey C., Foltz C., Hewett P. 1995, ApJ, 445, 62

   \bibitem{}
Jones L.R., McHardy I.M., Merrifield M.R., {\it et al.} 1996, MNRAS, in press
(astro-ph/9610124)

   \bibitem{}
K\"ohler T. 1996, PhD Thesis, University of Hamburgh 

   \bibitem{}
Kron, R.G. 1980, ApJ, 43, 305

   \bibitem{}
La Franca F., Cristiani S., Barbieri C. 1992, AJ, 103, 1062 

   \bibitem{}
La Franca F., Gregorini L., Cristiani S., de Ruiter H., Owen F. 1994, AJ,
108, 1548
 
   \bibitem{}
La Franca F., Franceschini A., Cristiani S., Vio R. 1995, A\&A, 299, 19

   \bibitem{}
Maccacaro T., Della Ceca R., Gioa I., Morris S., Stocke J., Wolter A. 1991,
ApJ, 374, 117

   \bibitem{}
Madau P., Ghisellini G., Fabian A.C. 1994, 270, L17

   \bibitem{}
Marshall H.L., Avni Y., Tananmbaum H, Zamorani G. 1983, ApJ, 269, 35

   \bibitem{}
Miller L., Peacock J.A., Mead A.R.G. 1990, MNRAS, 218, 265

   \bibitem{}
Miller L., Goldschmidt P., La Franca F., Cristiani F. 1993, in Observational
Cosmology, ASPC 51, eds. G. Chincarini, A. Iovino, T. Maccacaro and D.
Maccagni, p. 614

   \bibitem{}
Padovani P. 1993, 1993, MNRAS, 263, 461

   \bibitem{}
Press W.H., Teukolsky S.A., Vetterling W.T., Flannery B.P. 1992,
Numerical Recipes, Second Edition, Cambridge Univ. Press,
Cambridge, 640

   \bibitem{}
Sargent W.L.L.S., Steidel C.C., Boksenberg A. 1989, ApJS, 69, 703

   \bibitem{}
Schmidt, M., Green, R.F. 1983, ApJ, 269, 352

   \bibitem{}
Schneider D.P., van Gorkom J.H., Schmidt M., Gunn J.E. 1992, AJ, 103, 1451

   \bibitem{}
Schneider D.P., Schmidt M., Gunn J.E. 1991, ApJS, 101, 2004

   \bibitem{}
Visnovsky K.L., Impey C.D., Hewett P.C., Weymann R.J., Morris S.L. 1992,
ApJ, 391, 560

   \bibitem{}
Wampler E.J., Ponz D. 1985, ApJ, 298, 448

   \bibitem{}
Warren S.J., Hewett P.C., Osmer P.S. 1994, ApJ, 421, 412

   \bibitem{}
Wisotzki L., K\"ohler T., Groote D., Reimers D. 1996a, A\&AS, 115, 227

   \bibitem{}
Wisotzki L., Bade N., Engels D., Groote D., Hagen H.J., K\"ohler T., 
Reimers D. 1996b, in ``Wide Field Spectroscopy'', Athens, eds E.
Kontizas {\it et al.} 

   \bibitem{}
Zitelli, V., Mignoli, M., Zamorani, G., Marano, B., Boyle, B.J. 1992, MNRAS,
256, 349

\end{thebibliography}
\end{document}